\documentclass[prb, superscriptaddress, showpacs, floatfix, twocolumn]{revtex4}
\usepackage{amsmath}
\usepackage{bbm}
\usepackage{graphicx}
\usepackage{float}
\usepackage{epsfig}
\usepackage{epsf}
\usepackage{hyperref}

\usepackage{soul}
\usepackage{color}

\begin{document}
\title{Theory of ferromagnetism in vanadium-oxide based perovskites}
\author{Hung T. Dang}
\author{Andrew J. Millis}
\affiliation{Department of Physics, Columbia University, 538 West 120th Street, New York, New York 10027, USA}
\date{\today}

\begin{abstract}
The conditions under which ferromagnetism may occur in transition metal oxides with partially filled $t_{2g}$ shells such as vanadium-based perovskites are studied using a combination of density functional and single-site dynamical mean field methods. For reasonable values of the correlation strength, rotations of the VO$_6$ octahedra play an important role in enabling ferromagnetism, with ferromagnetism typically occurring for rotations larger than a nonzero critical value. Ferromagnetism is suppressed near the Mott-insulating phase but the phase boundary is otherwise only weakly dependent on carrier concentration. Design rules are suggested for new oxide systems  exhibiting ferromagnetism. 
\end{abstract}
% 71.28.+d      Narrow-band systems; intermediate-valence solids
% 75.10.Lp      Band and itinerant models

\maketitle

\section{Introduction\label{sec:intro}}
Transition metal oxides\cite{RevModPhys.70.1039} are of great interest in condensed matter physics because they exhibit a rich variety of exotic phenomena which remain incompletely understood. While ``late'' transition metal oxides (involving Cu or Ni) have been very extensively studied due to their connection to high-$T_c$ superconductivity and Mn-based compounds have attracted attention for their colossal magnetoresistance, the ``early'' transition metal oxides such as vanadium oxides have, despite some important studies, \cite{Pavarini04} received less  attention in recent years. However, following the pioneering work of Ohtomo and Hwang,\cite{Ohtomo_Hwang} early transition metal oxides are increasingly used as components of atomic-scale oxide-based  heterostructures. \cite{Kourkotis10,PhysRevB.80.241102}  One goal of heterostructure research is to design materials exhibiting phases not observed in bulk.\cite{Millis11,Hwang12} An essential step towards  realizing this goal  is obtaining  a clear understanding of the relationship between physical structure and observed electronic phenomena.

In this paper we investigate the relationship between lattice structure, correlation strength and electronic properties  in the context of ferromagnetism in early transition metal oxides. Ferromagnetism is a correlated electronic property which is both technologically important and (because only a spin symmetry and not translation or gauge symmetry is broken)  more straightforward from the theoretical and computational points of view than other phases such as antiferromagnetism or superconductivity.  We choose the early transition metal oxides in part because of the intriguing recent report\cite{PhysRevB.80.241102}  of ferromagnetism in superlattice systems involving LaVO$_3$ and SrVO$_3$. The report is of interest because ferromagnetism is reported for the superlattice  even though no ferromagnetism is observed in bulk solid solutions of the form La$_{1-x}$Sr$_x$VO$_3$. One possible explanation is that the superlattice  enables a different crystal structure, more favorable to ferromagnetism than that found in the observed bulk structures. Understanding whether this explanation is viable, and more generally being able to design superlattices with desired magnetic properties, requires deeper insight into the conditions 

The conditions under which ferromagnetism may occur is a question of  long-standing theoretical interest.\cite{Stoner,PhysRev.147.392,PhysRevB.41.2375,J.Low.Temp.Phys.99.349} The development of dynamical mean field theory \cite{Georges96} has opened a new avenue of research, but apart from some pioneering investigations of the Curie temperatures of Fe and Ni \cite{PhysRevLett.87.067205} the studies have mainly been based on model systems. Vollhardt, Ulmke, and collaborators have studied the single-band Hubbard model, finding that in this model ferromagnetism occurs at generic carrier concentrations only when there is a large density of states peak at or near the lower band edge.\cite{springerlink:10.1007/s002570050375,Euro.Phys.B1.301} For a fixed value of  the Hubbard $U$, the  Curie temperature $T_c$ was found to  depend sensitively on the peak position,  becoming unmeasurably small as the density of states peak was moved a small distance away from the lower band edge.\cite{PhysRevB.58.12749} However, many materials of physical and technological interest involve transition 
metals with partially filled degenerate (or nearly degenerate) $d$ levels, where the Hund's interaction may play an important role. While the importance of the Hund's interaction in partially filled $d$ levels has been appreciated for decades, the issue has  been systematically studied only in the case of the Bethe lattice \cite{Euro.Phys.B5.473,PhysRevB.83.125110,PhysRevB.80.235114} in which the density of states has a simple semicircular structure. In this situation large values of the Hund's coupling and correlation strength (the Hubbard $U$) are required for ferromagnetism. 

\begin{figure}
 \includegraphics[width=0.67\columnwidth]{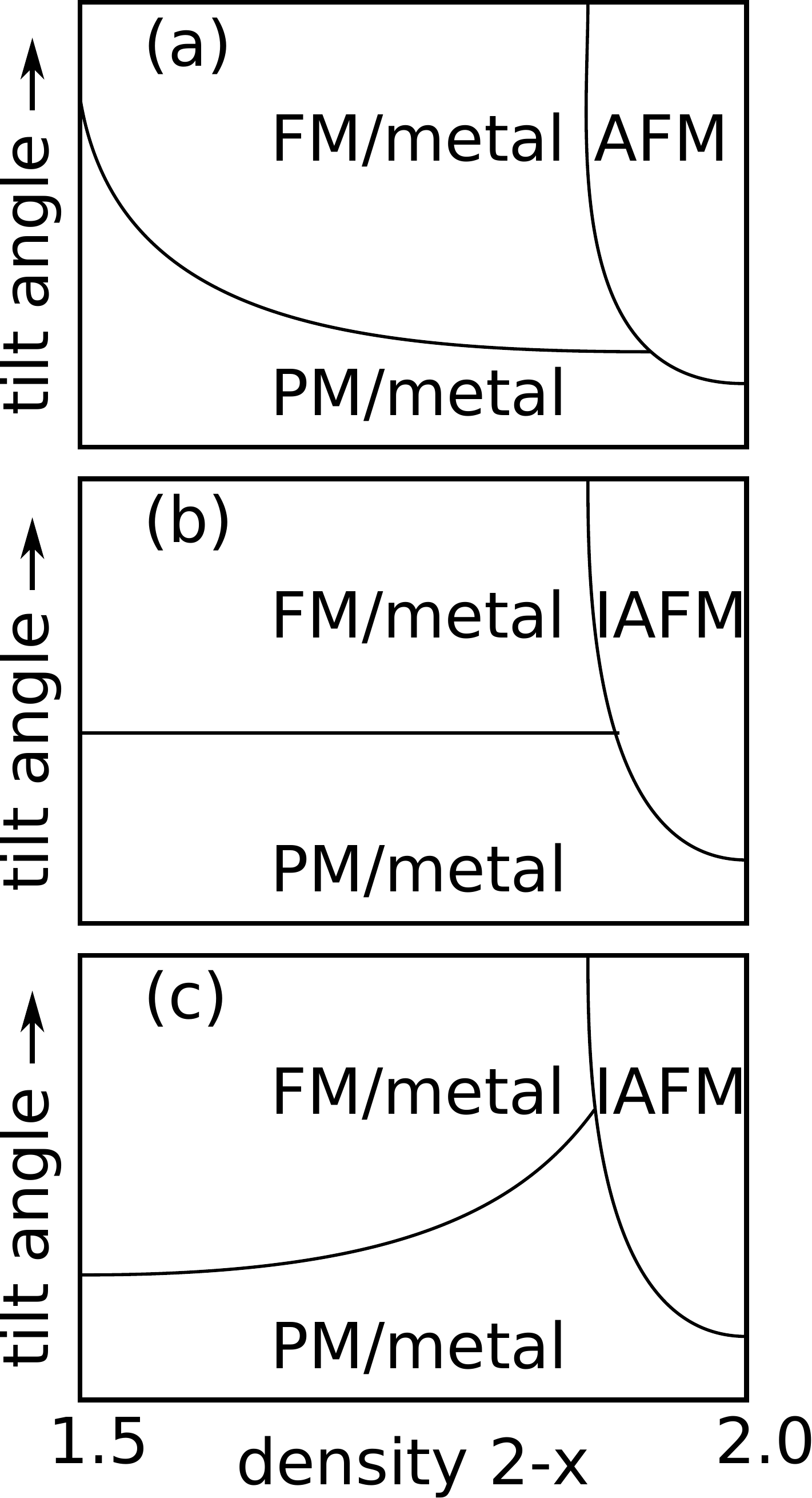}
\caption{\label{fig:schematic_pd} Three possible magnetic phase diagrams for bulk solid solution La$_{1-x}$Sr$_x$VO$_3$ as a function of carrier density ($x$ axis) and tilt angle ($y$ axis). The phase boundaries are the black curves. (a) Weak correlations: Slater antiferromagnet at $x=0$. (b) Intermediate correlations. (c) Strong correlations: Mott insulator at $x=0$. The notations are: FM for ferromagnetic, PM for paramagnetic, AFM for antiferromagnetic order, and I for insulating state.}
\end{figure}

In this paper, we study the conditions under which ferromagnetism may occur in systems related to  La$_{1-x}$Sr$_x$VO$_3$ (LSVO). In this family of materials the key orbitals are vanadium $d$ levels. These are partly filled (in La$_{1-x}$Sr$_x$VO$_3$ the mean number of $d$ electrons is $2-x$) and the electrons are subject to a strong Hund's coupling, favoring magnetism. The end-member LaVO$_3$ is classified as a Mott insulator. \cite{RevModPhys.70.1039} It has  structural and antiferromagnetic transitions at around 140K but is insulating to high temperatures.\cite{Bordet1993253,PhysRevLett.85.5388} Our investigations are based on density functional theory (DFT) calculations and include  in particular  the effects of GdFeO$_3$-type rotations of the BO$_6$ structural motif of the ideal perovskite structure. We treat the many body physics using the single-site dynamical mean field approximation,\cite{Georges96} which is widely used in real-materials many-body physics calculations\cite{Kotliar06,Held07} and in particular in studies of ferromagnetism. \cite{Vollhardt.Review,springerlink:10.1007/s002570050375}

Our main result is an understanding of the dependence of ferromagnetism on   carrier concentration, octahedral rotation and correlation strength. At non-integer electron concentration we find that increasing the octahedral rotation amplitude increases the tendency to ferromagnetism.  The physics underlying this result is a density of states effect related to that previously noted in the one-band Hubbard model.\cite{Vollhardt.Review,springerlink:10.1007/s002570050375} We also find that at fixed rotation amplitude the carrier concentration dependence of ferromagnetism depends on the correlation strength.  For weak to moderate correlations (so that  at the carrier concentration $n=2$ corresponding to LaVO$_3$ the paramagnetic phase is metallic in contradiction to experiment \cite{RevModPhys.70.1039,Bordet1993253,PhysRevLett.85.5388}) decreasing the carrier concentration below $n=2$ weakly decreases the tendency to ferromagnetism. For stronger correlations (so that  at the carrier concentration $n=2$ corresponding to LaVO$_3$, the insulating phase is a Mott insulator), doping away from LaVO$_3$ strongly increases the tendency towards ferromagnetism. An intermediate situation may also occur.  The three panels of Fig.~\ref{fig:schematic_pd} present schematic representations of these three cases. 

We remark that in the literature, the many-body physics properties of transition metal oxides have been  modeled in two ways. One, which we refer to as the ``$d$ only model,'' treats the frontier orbitals (in the case of the vanadium-based materials, antibonding bands derived mainly from V $t_{2g}$-symmetry $d$ orbitals admixed with oxygen $p_\pi$ states) as a multiorbital Hubbard model. This approach is generally accepted \cite{RevModPhys.70.1039} as the relevant description of the early transition metal perovskites such as the systems we study here. A second approach is based on atomic-like $d$ orbitals which are subject to on-site correlations and are hybridized with the full $p$ manifold. This approach is used in the DFT+U and DFT+DMFT \cite{Kotliar06,Held07} approaches. It is generally believed to be essential to a correct description of the late (e.g. Cu-based or Ni-based)  transition metal oxides \cite{PhysRevLett.55.418} but has been less widely used in the study of early transition metal oxides. We consider both approaches in this paper, finding similar qualitative conclusions but significant quantitative differences which arise from the richer physics of the ``$pd$ model'' situation. 

The paper has following structure. The model and methods  are described in Sec.~\ref{sec:model}. Section~\ref{sec:bulkphasediagram} derives the magnetic phase diagram of LaVO$_3$ as a function of carrier concentration, tilt angle and correlation strength using the widely used $d$ only model. In this analysis, most attention is given to  the strongly correlated case presented in  Fig.~\ref{fig:schematic_pd}c. Section~\ref{sec:pdmodel} analyzes the $pd$ model. The final Sec.~\ref{sec:conclusions} is the summary and conclusion. Appendices present details of some of the results.

\section{Model and methods\label{sec:model}}
\subsection{Model}
\begin{figure}
 \includegraphics[width=0.70\columnwidth]{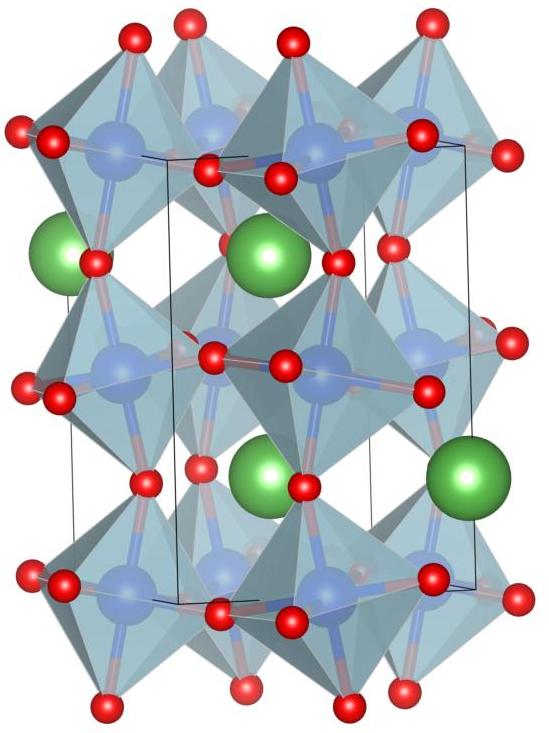}
\caption{\label{fig:GdFeO3}(Color online) Representation of $AB$O$_3$ perovskite structure with GdFeO$_3$-type (octahedral rotation) distortion generated using the VESTA program.\cite{Momma:db5098} Large spheres (green online): A atoms (here La or Sr), small spheres (red online): O atoms, intermediate size spheres at the centers of the octahedra  (blue online) : B-site atoms (here, V).}
\end{figure}

We study systems derived from SrVO$_3$ and LaVO$_3$. Bulk SrVO$_3$ crystallizes in the ideal cubic ABO$_3$ perovskite structure.\cite{Rey1990101} The crystal structure of  LaVO$_3$ may be thought of as a cubic perovskite with an additional GdFeO$_3$-type rotational distortion leading to a tilted structure with symmetry group $Pnma$\cite{Bordet1993253} Fig.~\ref{fig:GdFeO3} shows the lattice structure of LaVO$_3$. The basic structural motif of the perovskite structure is the oxygen octahedron. In the GdFeO$_3$ rotated structure there are four inequivalent octahedra characterized by different directions of the principal axes and by slight ($\sim 5\%$) distortions of the V-O bond length. We have found (not shown) that varying the relative magnitudes of the V-O bond lengths over the physical range causes only small changes to the non-interacting DOS or to the DMFT solution. We therefore set all V-O distances to $d=1.95$\AA~ and focus on the effect of rotation by studying a range of $\theta$ and $\phi$.

The octahedral rotations in the GdFeO$_3$-distorted  $Pnma$ systems can be characterized by two angles, $\theta$ and $\phi$,\cite{1367-2630-7-1-188} with corresponding rotation axes $\hat{n}_\theta$, $\hat{n}_\phi$ and wavevectors $\vec{Q}_\theta$, $\vec{Q}_\phi$ characterizing the changes in rotation axis from cell to cell. We choose coordinates such that the rotation axis $\hat{n}_\theta$ is $[110]$ while $\hat{n}_\phi$ is $[001]$. The corresponding wavevectors are $\vec{Q}_\theta=(\pi,\pi,\pi)$ and $\vec{Q}_\phi=(\pi,\pi,0)$. For bulk LaVO$_3$, $\theta=11.5^\circ$ and $\phi=8.8^\circ$.\cite{Bordet1993253} As La is progressively replaced by Sr in bulk solid solution LSVO, $\theta$ and $\phi$ go to $0$. Appendix~A provides an estimate of the $x$-dependence of $\theta$ and $\phi$ in the actual solid solution. 

We now turn to the electronic structure. The large energy splitting between the transition metal $d$ bands and oxygen $p$ bands characteristic of early transition metal oxides such as La/SrVO$_3$ is generally believed\cite{RevModPhys.70.1039} to justify a downfolding to a model containing only the frontier bands, which in the present case are composed mainly of V-derived $t_{2g}$ symmetry $d$ states with only a small admixture of oxygen $p$ states. We refer to this as the $d$ only model. In a subsequent section we present an  examination of a more general ``charge transfer'' model in which the full $p$ $d$ complex is considered.

The Hamiltonian of the $d$ only model is
\begin{equation}
 H = H_{kin} + H_{onsite},
\label{eqn:fullH}
\end{equation}
where $H_{kin}$ describes the electron hopping between sites and $H_{onsite}$ describes the $d$ $d$ interactions, which we take to be on-site.

The kinetic Hamiltonian has the quadratic form 
\begin{equation}\label{eqn:Hkin}
H_{kin} = \sum_{\mathbf{k},\alpha,\beta,\sigma} H_{band}^{\alpha\beta}(\mathbf{k})c^\dagger_{\mathbf{k}\alpha\sigma} c_{\mathbf{k}\beta\sigma},
\end{equation}
where $c^\dagger_{\mathbf{k}\alpha\sigma}$ and $c_{\mathbf{k}\beta\sigma}$ are electron creation and annihilation operators in reciprocal space with wavevector $\mathbf{k}$. $\alpha$ and $\beta$ are orbital indices, and $\sigma$ is the spin index. 

For the cubic structure, the  $\hat{H}_{band}(\mathbf{k})$ describing the $t_{2g}$-derived antibonding bands is easily constructed using phenomenological tight binding methods.\cite{PhysRev.94.1498} For example, for calculations with only antibonding bands composed mainly of $t_{2g}$ orbitals, $\hat{H}_{band}(\mathbf{k})$ is almost diagonal and well approximated by a two-dimensional energy dispersion with nearest neighbor hopping $t$ and next-nearest neighbor hopping $t'$. The explicit form for the band arising from the $xy$-symmetry orbital is
\begin{equation}\label{eqn:dispersion2D}
\epsilon(\mathbf{k})_{xy} = -2t(\cos k_x+\cos k_y) - 4t'\cos k_x\cos k_y.
\end{equation}
and for the $xz/yz$-derived bands one relabels the momentum indices appropriately. 

For the tilted structure, the form of $\hat{H}_{band}(\mathbf{k})$ is more difficult to express in the simple tight binding language used for the cubic perovskite structure. First, the natural basis for defining local $d$ orbitals is aligned with the axes of the local VO$_6$ octahedron and so varies from site to site making the Slater-Kanamori procedure much more involved. Second, the V-O-V bond is buckled, decreasing the hopping $t$ both in absolute terms and relative to $t'$, as well as opening many other channels of hopping from one V-site to another. For tilted structures we construct $\hat{H}_{band}(\mathbf{k})$ numerically from density functional theory (DFT) band structure calculations using  maximally-localized Wannier function (MLWF) techniques.\cite{PhysRevB.56.12847} We have verified that when applied to the cubic structure the method reproduces the phenomenological tight binding model discussed above.

We found that in the resulting MLWF band structures (not shown), the bands derived from $t_{2g}$ and oxygen $p$ orbitals agree very well with the DFT results, but the $e_g$ antibonding bands do not because they are entangled with many other bands. However, the $e_g$ states derived from these bands are well above  the Fermi level and, in an appropriate local basis aligned with the symmetry axes of a VO$_6$ octahedron, do not mix with $t_{2g}$ bands. They therefore do not affect the final results significantly.

The on-site interaction term in the Hamiltonian, $H_{onsite}$, is assumed to take the standard Slater-Kanamori form \cite{PhysRev.49.537,Kanamori195987,PTP.30.275}
\begin{equation}\label{eqn:Honsite}
\begin{split}
H_{onsite} & = U\sum_{i\alpha}n_{i\alpha\uparrow}n_{i\alpha\downarrow}  + U_1\sum_{i\alpha\neq \beta}n_{i\alpha\uparrow}n_{j\beta\downarrow} + \\
	& + U_2\sum_{i,\alpha > \beta,\sigma}n_{i\alpha\sigma}n_{i\beta\sigma} + \\
        & + J_{ex}\sum_{i\alpha\neq \beta}\psi^\dagger_{i\alpha\uparrow}\psi_{i\beta\uparrow}\psi^\dagger_{i\beta\downarrow}\psi_{i\alpha\downarrow} + \\
	& + J_{pair}\sum_{i\alpha\neq \beta}\psi^\dagger_{i\alpha\uparrow}\psi_{i\beta\uparrow}\psi^\dagger_{i\alpha\downarrow}\psi_{i\beta\downarrow},
\end{split}
\end{equation}
where $U$ is the on-site intra-orbital interaction, $U_2$ and $U_1$ are on-site inter-orbital interactions for electrons with the same and different spin respectively and $J_{ex}$ and $J_{pair}$ are the  exchange and pair hopping parts of the interaction. For a transition metal ion in free space the relations between these coefficients are: $J_{ex} = J_{pair} = J$, $U_1=U-2J$, $U_2 = U_1-J=U-3J$.\cite{PhysRevB.56.12909} It is generally believed that the relations also hold for the $t_{2g}$ shell\cite{RevModPhys.70.1039} and we make this assumption here. We consider several values of $U$ and $J$ but focus most attention on the values  $U=6\mathrm{eV}\sim22t$ and $J=1$eV. These values of $U$ and $J$ are similar to but  slightly larger than those used in recent papers\cite{PhysRevB.73.155112,PhysRevLett.93.156402,PhysRevLett.99.126402,EPL99.67003} and are chosen to reproduce the crucial feature of the phase diagram, which is that LaVO$_3$ is a Mott insulator.

\subsection{Dynamical mean-field theory}
To treat the on-site interaction terms we use  single-site dynamical mean field theory (DMFT),\cite{Georges96} which obtains the electron self-energy $\hat{\Sigma}(\omega)$  (assumed to be site-local but dependent on spin and orbital indices) from the solution of a quantum impurity model.  We use the hybridization-expansion CTQMC solver\cite{PhysRevLett.97.076405} to solve the quantum impurity model. In the present case, the impurity model is a three-orbital model specified by the $U$ and $J$ interactions and by a   hybridization function fixed by the self-consistency condition
\begin{equation}
 \hat{G}^i_{imp}(\omega) = \left\{\left[(\omega+\mu)\textbf{1} - \hat{H}_{band} - \hat{\Sigma}(\omega) \right]^{-1}\right\}_{ii}.
\end{equation}
where $\{\}_{ii}$ denotes the on-site Green's function at vanadium site $i$.

In applying the procedure to the GdFeO$_3$-distorted structure, it is important to formulate the impurity problem corresponding to a given V site in a local basis aligned to the V-O bond directions of the VO$_6$ octahedron at this site.  The self energy derived from solving the impurity problem for each octahedron is then rotated back to the global basis to define the lattice Green's function used in the self-consistency condition. There are three advantages. First, the hybridization function is nearly diagonal in the local basis, which greatly reduces the sign problem in the quantum Monte Carlo solution of the impurity model. Second, the $t_{2g}$ and $e_g$ bands are clearly distinguished and well separated in the local basis, enabling removal of the $e_g$ bands from the many-body problem (they can be neglected or treated in mean field) so the dynamical impurity model involves three rather than five bands. Third, the self-consistency conditions corresponding to the four inequivalent V-sites of $Pnma$ LaVO$_3$ are related by rotations so only one impurity model needs to be solved.

\subsection{Determining the magnetic phase boundary}
In order to detect magnetic order, one can allow for broken symmetry states in the DMFT procedure and lower the temperature until an ordered state is reached. However, ordering temperatures are typically so low relative to the basic energy scales in the problem that our simulations become prohibitively expensive. We therefore compute the susceptibility which is found to have a Curie-Weiss form $\chi^{-1}\sim T-T_c$. Extrapolation of $\chi^{-1}$ to $0$ yields an estimate of the Curie temperature $T_c$. We interpret positive $T_c$ as evidence of magnetism. To compute $\chi^{-1}$, we add a uniform field $\vec{H}\cdot\vec{\sigma}$ to our Hamiltonian, compute the magnetization $m(H)$, verify that $m$ is linear in $H$ and then define $\chi^{-1}=\dfrac{H}{m}$. 

\begin{figure}[h]
 \includegraphics[width=0.9\columnwidth]{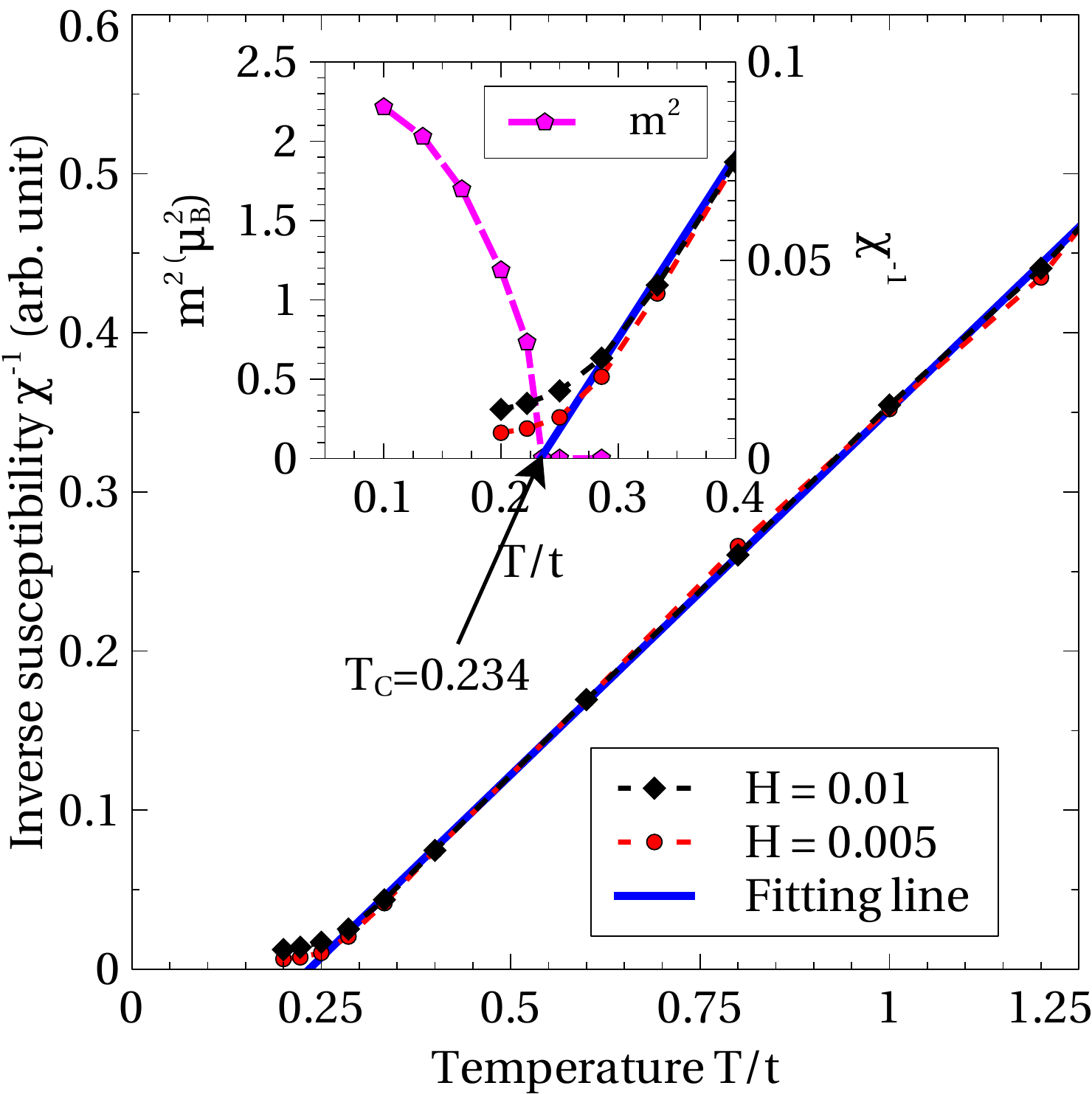}
\caption{\label{fig:demo_extrapolateTc}(Color online) Inverse magnetic susceptibility $\chi^{-1} = \dfrac{H}{m(H)}$ plotted vs. temperature $T$ obtained from single-site DMFT solution to Eqs.~(\ref{eqn:fullH} and \ref{eqn:dispersion2D}) with $H_{kin}$ taken to have the form appropriate to a simple cubic lattice with nearest neighbor hopping $t$ and second neighbor hopping $t'=-0.3t$. Carrier density is fixed to $n=1.5$ and applied magnetic field $H=0.005$ (circles) and $H=0.01$ (diamonds). The Curie temperature $T_c=0.234t$ is estimated by linear extrapolation of $\chi^{-1}$ in the region where $m$ is linear in $H$. Inset: Expanded view of the near-$T_c$ region, together with the magnetization squared ($m^2$) (left $y$ axis) calculated by allowing for broken symmetry DMFT solution at $H=0$ which also shows that $T_c\approx0.222t$. The parameters are $t=0.5,t'=-0.3t,U=16=32t$ and $J=\dfrac{U}{6}$.}
\end{figure} 

Fig.~\ref{fig:demo_extrapolateTc} demonstrates the method on a model with a high Curie temperature, where the magnetic state can easily be constructed.  This  model is defined by Eqs.~(\ref{eqn:fullH},~\ref{eqn:dispersion2D})  with $H_{kin}$ taking the same form as in the simple cubic lattice but with $t'$ chosen as $\dfrac{t'}{t}=-0.3$, so that  the sign of $t^{'}/t$ is opposite to the sign implied by band structure calculations on the actual materials. For this unphysical sign of $t^{'}/t$   the model has a ferromagnetic ground state with a high Curie temperature.  We set  carrier density $n=1.5$ and calculate $T_c$  in two ways: by lowering the temperature until ferromagnetic order is observed, which is at $T_c\approx0.222t$; or by measuring $\chi^{-1}(T)$ for several values of $T$ above $T_c$ and linearly extrapolating to $\chi^{-1}=0$. The extrapolation shows $T_c=0.234t$. The two values are very close. We conclude that extrapolating $T_c$ from $\chi^{-1}(T)$ is a reliable way to determine whether the model exhibits ferromagnetic order.

\section{Magnetic phase diagram\label{sec:bulkphasediagram}}
In this section, we study the magnetic phase diagram of model systems derived from the calculated band structure of LaVO$_3$ but  with variable amplitude for the  GdFeO$_3$ distortion and different values for the carrier concentration.

\begin{figure}[h]
 \includegraphics[width=0.9\columnwidth]{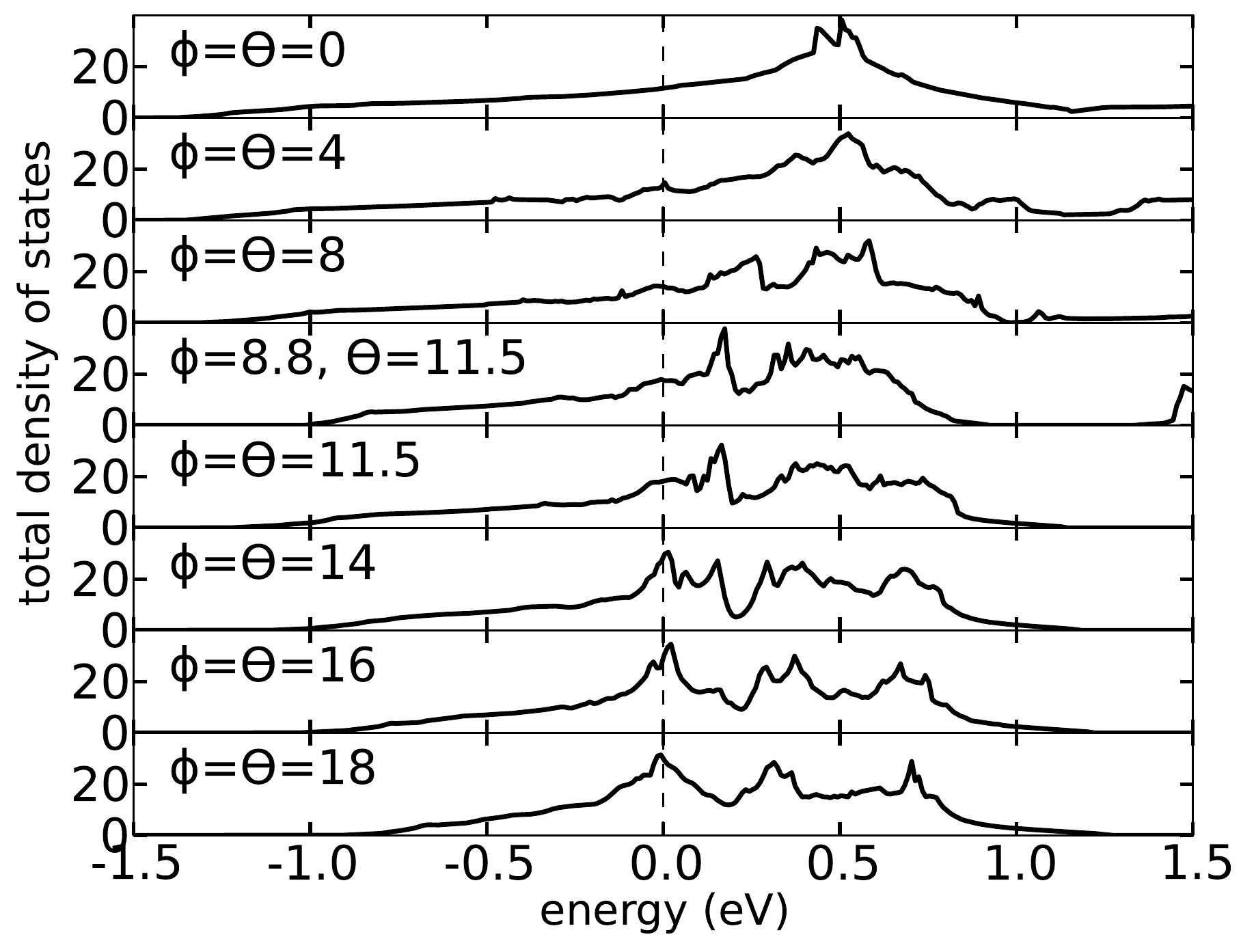}
\caption{\label{fig:DOS_bulk} Density of states for bulk LaVO$_3$ ($d$ carrier density $n=2$) calculated from DFT with different values of tilt angle $\theta$ and rotation angle $\phi$ in which $\theta=\phi$ together with the DOS of realistic LaVO$_3$ structure ($\theta=11.5^\circ,\phi=8.8^\circ$). A van Hove peak at the Fermi level develops as $\theta$ and $\phi$ increase, and is well-formed at $\theta=\phi=14^\circ$ and above. The bandwidth $W$ decreases as the $\theta$ and $\phi$ increases. The Fermi energy is at $0$.}
\end{figure}

\begin{figure}[h]
 \includegraphics[width=0.9\columnwidth]{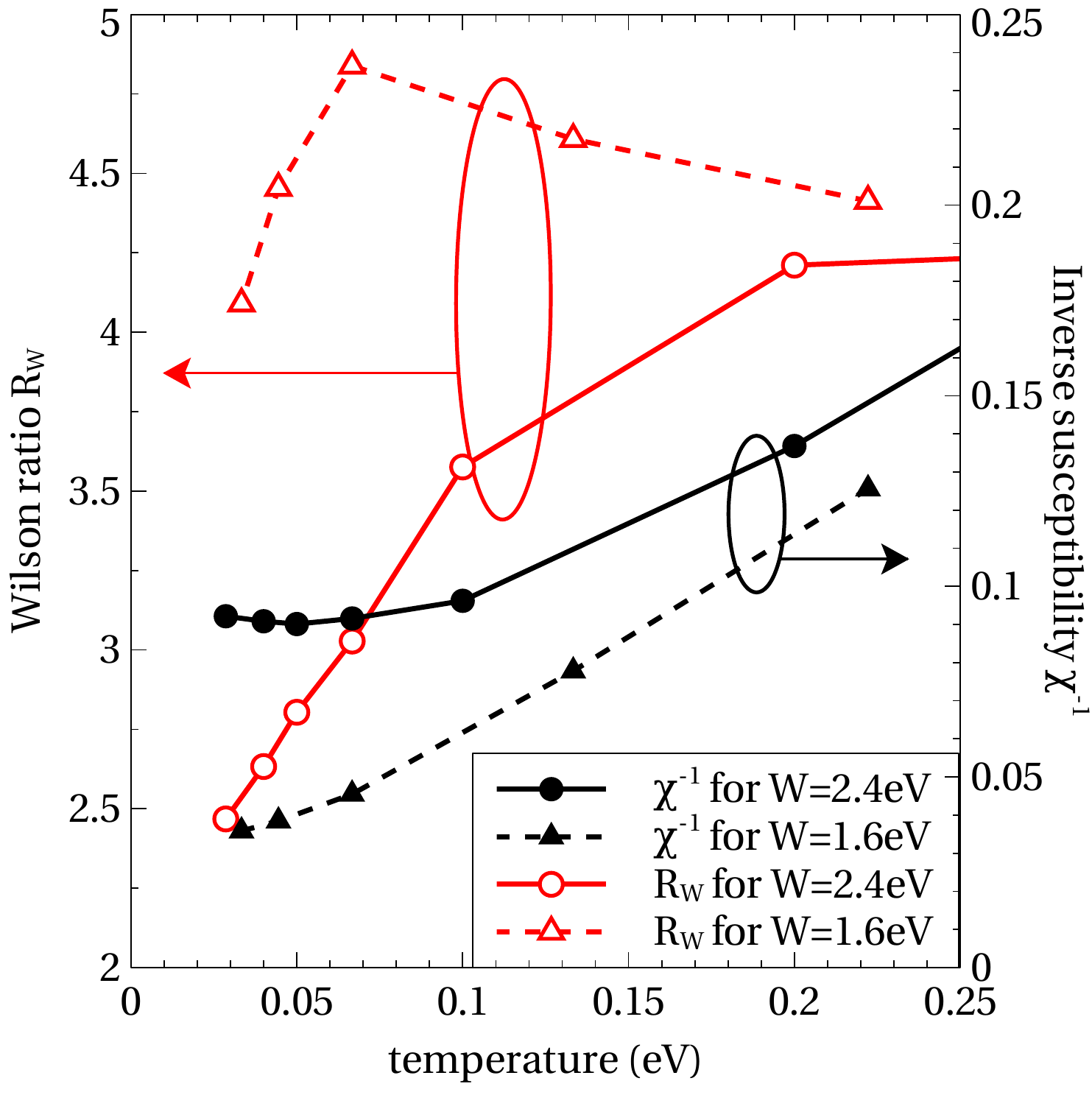}
\caption{\label{fig:cubic_Wilson}(Color online) Inverse magnetic susceptibility $\chi^{-1}$ (closed symbols, black on-line; right axis) and Wilson ratio $R_W$ (open symbols, red on-line, left axis)  calculated for La$_{1-x}$Sr$_x$VO$_3$ cubic structure as functions of temperature at carrier density  $n=1.5$ and on-site interactions  $U=6$eV, $J=1$eV using two bandwidths, $W=2.4$eV (circles and solid lines) and $W=1.6$eV (triangles and dashed lines).}
\end{figure} 

Fig.~\ref{fig:DOS_bulk} shows the evolution of the non-interacting density of states with increasing amplitude of the GdFeO$_3$ distortion. If the structure is cubic, the energy dispersion in the $t_{2g}$ manifold is almost two dimensional. The sign and magnitude of $t'$ place the van Hove peaks of the non-interacting DOS at high energy (see top panel ($\theta=\phi=0$) of Fig.~\ref{fig:DOS_bulk}). As the distortion amplitude is increased the lowest DOS peak shifts to lower energy and the bandwidth decreases. Previous literature suggests a connection between peak position and ferromagnetism;\cite{Vollhardt.Review,springerlink:10.1007/s002570050375} while standard Stoner theory\cite{Stoner} suggests that bandwidth may also be important. 

We begin our study of the connection by examining the cubic structure. The solid  curve with filled circles (black online) in Fig.~\ref{fig:cubic_Wilson} is the inverse susceptibility calculated for temperatures down to $T=0.03$eV for the case $U=6$eV, $J=1$eV and carrier density $n=1.5$ using the DOS shown in top panel of Fig.~\ref{fig:DOS_bulk}, which  has the bandwidth $W\approx2.4$eV. The  inverse susceptibility curve is seen to deviate from the Curie form at low temperature. The extrapolation to zero of the high temperature linear regime implies a  $T_c<0$.

We attribute the flattening out of the susceptibility curve to the onset of Fermi-liquid coherence. To verify this, and to gain additional insight into the nature of spin correlations in this model, we computed the Wilson ratio $R_W=\dfrac{\pi^2 }{3}\dfrac{\chi}{\gamma}$,\cite{RevModPhys.47.773} (in our conventions  the dimensional factors $\dfrac{k_B}{\mu_B}=1$). Here $\gamma$ is the coefficient of the linear specific heat $\gamma=\left.\dfrac{dC_V}{dT}\right|_{T=0}=\dfrac{\pi^2}{3}\mathrm{Tr}[\nu_F Z^{-1}]$.\cite{PhysRevB.75.235107} We estimated the linear coefficient of the specific heat from the density of states $\nu_F$ at the Fermi level and calculated renormalization factor $Z$, which are obtained from the measured imaginary time Green's function via  $\nu_F \approx -\dfrac{\beta}{\pi} G(\tau=\beta/2)$ and from the Matsubara frequency self energy via $Z^{-1} \approx 1-\dfrac{\mathrm{Im}\Sigma(\omega_1=3\pi/\beta)-\mathrm{Im}\Sigma(\omega_0=\pi/\beta)}{\omega_1 - \omega_0}$ (not shown). 

The solid curve with open circles (red online) in Fig.~\ref{fig:cubic_Wilson} shows that, for the DOS in the top panel of Fig.~\ref{fig:DOS_bulk} with $U=6$eV and $J=1$eV, the Wilson ratio extrapolates to the value $R_W=2$ as temperature $T\rightarrow 0$.  The value $R_W=2$ is expected for a Kondo lattice with a low quasiparticle coherence scale but no intersite correlations,  while a system with strong ferromagnetic correlations would be characterized by an $R_W\gg2$. We therefore conclude that there is no evidence for ferromagnetism in the cubic structure at $U=6, J=1$eV and $W=2.4$eV and that the flattening of the $\chi^{-1}(T)$ curve indicates the onset of the Fermi liquid coherence.

We now turn to the effects of the GdFeO$_3$ distortion. For simplicity of presentation, we focus mainly on the case $\theta=\phi$. Fig.~\ref{fig:DOS_bulk} shows the evolution of DOS with tilting angle.  We see that, as the tilt angle is increased, the position in energy of the lowest density of states peak shifts down in energy.

\begin{figure}[h]
 \includegraphics[width=0.9\columnwidth]{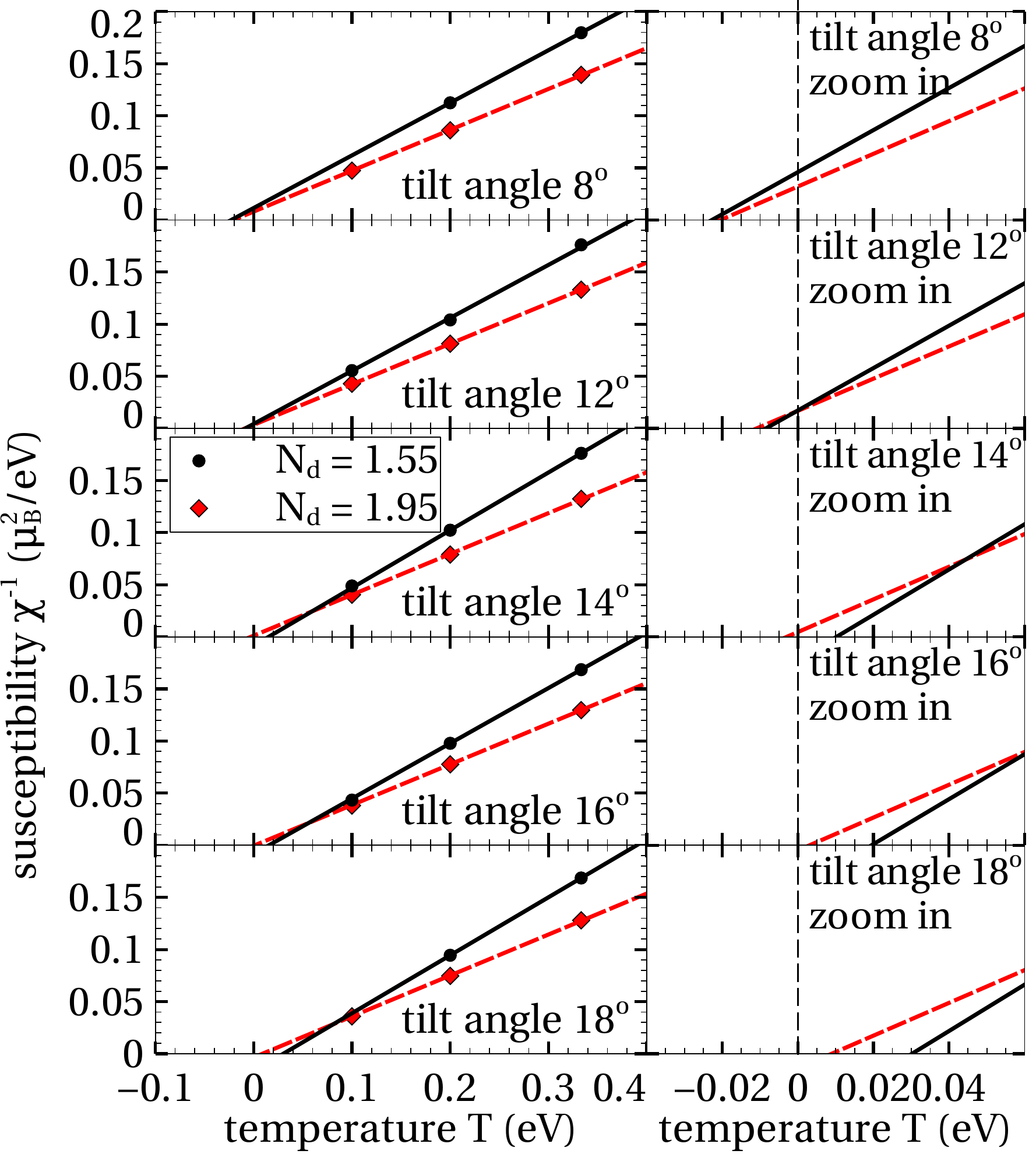}
\caption{\label{fig:Tc_trend_donly}(Color online) Temperature dependence of inverse susceptibility computed for model La$_{1-x}$Sr$_x$VO$_3$ computed for tilt angles $\theta=\phi$ from $8^\circ$ to $18^\circ$ computed at densities $n=1.55$ (black circle lines) and $n=1.95$ (diamond dashed lines) and  interaction parameters $U=6$eV, $J=1$eV from Eq.~\eqref{eqn:fullH} with $H_{kin}$ derived from MLWF fits to band structure. The diamonds and circles are data points measured by DMFT. The lines are fitted from data points. Left column: plot over wide temperature range; right column: expanded view of small $\chi^{-1}$ region. The vertical dashed line marks zero temperature.}
\end{figure} 

Fig.~\ref{fig:Tc_trend_donly} shows the evolution of the inverse susceptibility with increasing amplitude of GdFeO$_3$ distortion at $U=6$eV and $J=1$eV for two  carrier density values $n=1.55$ and $n=1.95$. For $n=1.55$, ferromagnetic order can be observed starting at $\theta=\phi=14^\circ$.  In contrast, for $n=1.95$, the Curie temperature is nonzero within errors  only for $\theta=\phi>16^\circ$.  

\begin{figure}[h]
 \includegraphics[width=0.9\columnwidth]{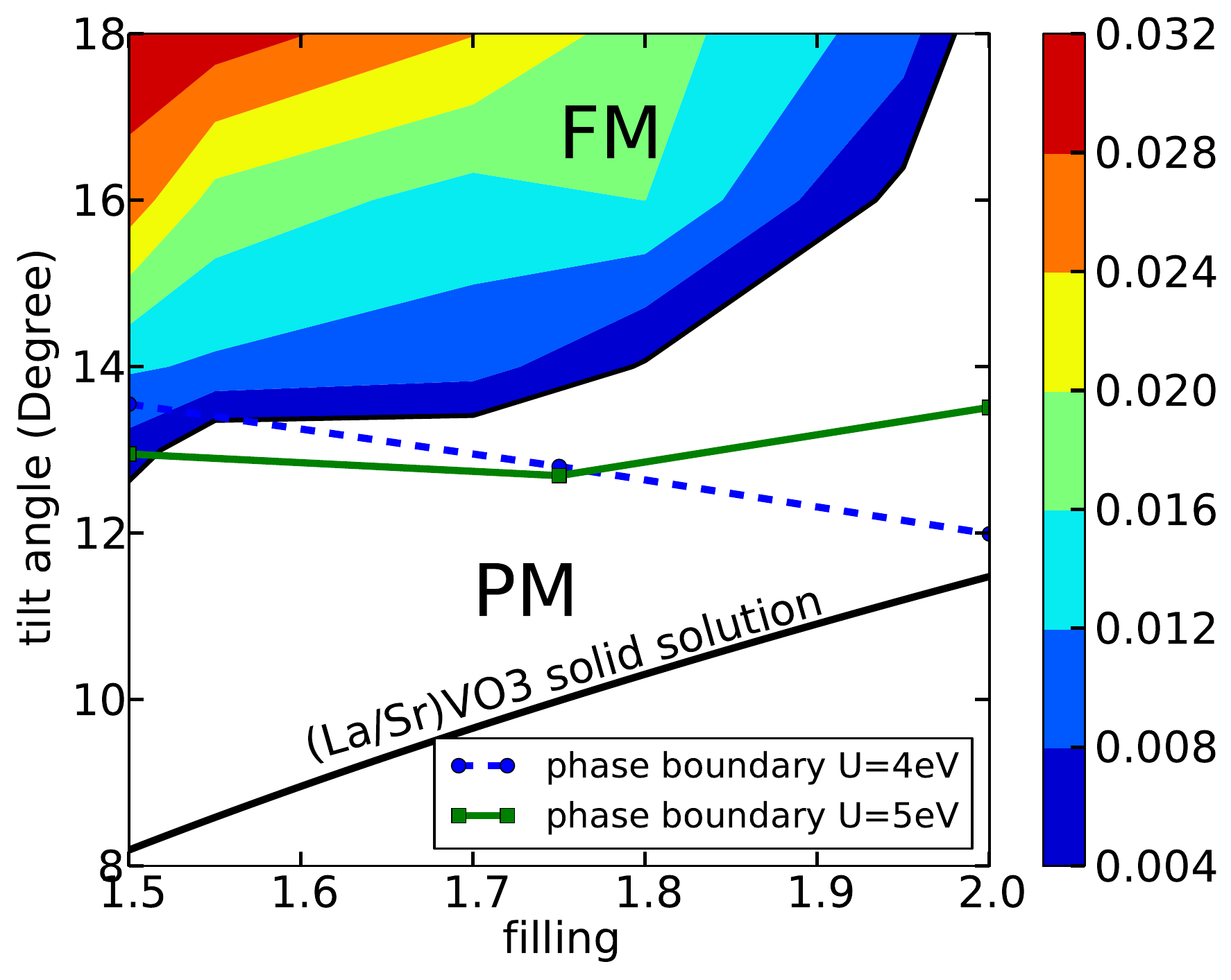}
\caption{\label{fig:phase_diagram}(Color online) The magnetic phase diagram in the space of  carrier density $n$ ($x$-axis) and  tilt and rotation angles $\theta=\phi$ ( $y$-axis) for the solid solution La$_{1-x}$Sr$_x$VO$_3$ with on-site interactions $U=6$eV, $J=1$eV. The white region indicates paramagnetic order ($T_c < 0.004$eV), the colored regions denote ferromagnetic order with the transition temperature $T_c$ given by the scale bar (color on-line). The solid curve is the $\theta = \theta(n)$ curve estimated for La$_{1-x}$Sr$_x$VO$_3$ as described in Appendix A. The estimated phase boundaries for $U=4$ (circle and dashed line, blue on-line) and $5$eV (squares and  solid line, green on-line) with $J=1$eV are also plotted.}
\end{figure} 

The shaded areas (color online) in Fig.~\ref{fig:phase_diagram} show the phase diagram resulting from a detailed study of  the dependence of Curie temperature on carrier density and tilt (rotation) angle at $U=6$eV and $J=1$eV obtained from curves such as those shown in Fig.~\ref{fig:Tc_trend_donly}. The expense of the computation and the uncertainties inherent in our extrapolation means that the phase boundary is not precisely determined. We regard $T_c<0.004$eV as consistent with $T_c=0$ within uncertainties. The width of the strip separating $T_c=0.004$ and $0.008$ gives a measure of the error bars on the locations of the phase boundary. The divergence in tilt angle required to obtain a nonzero transition temperature as $n\rightarrow 2$ may be understood from the fact that for the correlation strength considered here the $n=2$ compound is a Mott insulator. Ferromagnetism is favored by metallic motion of the carriers and is suppressed in proximity to the Mott insulating phase which for the parameters we consider is antiferromagnetically ordered.  To summarize, we see that for  these values of $U$ and $J$  obtaining a ferromagnetic state requires two conditions: (1) large hole doping away from LaVO$_3$ and (2) large GeFeO$_3$-type distortion.

We now consider the physics behind the dependence of transition temperature on tilt angle. Inspection of Fig.~\ref{fig:DOS_bulk} shows that increasing the magnitude of the  GdFeO$_3$ both reduces the bandwidth (from $2.4$eV at $\theta=\phi=0$ to $1.8$eV at the critical angle) and moves the position of the lowest density of states peak to lower energy. To investigate the relative importance of the two  effects we have performed computations for the cubic structure using $U=6$eV, $J=1$eV   but with reduced bandwidth $W=1.6$ eV. The dashed curves  in Fig.~\ref{fig:cubic_Wilson} show the Wilson ratio $R_W$ (open triangles, red on-line) and the inverse susceptibility $\chi^{-1}$ (filled triangles, black on-line) for the smaller bandwidth  $W=1.6$eV.  The susceptibility curve indicates that the intercept is increased relative to the larger bandwidth case, but remains negative. The sharp downturn in the Wilson ratio at the lowest temperatures suggests that the decrease in intercept is a consequence of a lowered  Fermi liquid coherence scale and does not indicate stronger ferromagnetic correlations. We therefore conclude that for reasonable values of the bandwidth and correlation strength the crucial factor for ferromagnetism is the position of the density of states peak. For very large values of the Hund's coupling $J(\gtrsim 3\mathrm{eV})$, ferromagnetism may occur over a much wider parameter range, as seen in previous Bethe lattice studies.\cite{Euro.Phys.B5.473,PhysRevB.83.125110,PhysRevB.80.235114} 

\begin{figure}[h]
 \includegraphics[width=0.9\columnwidth]{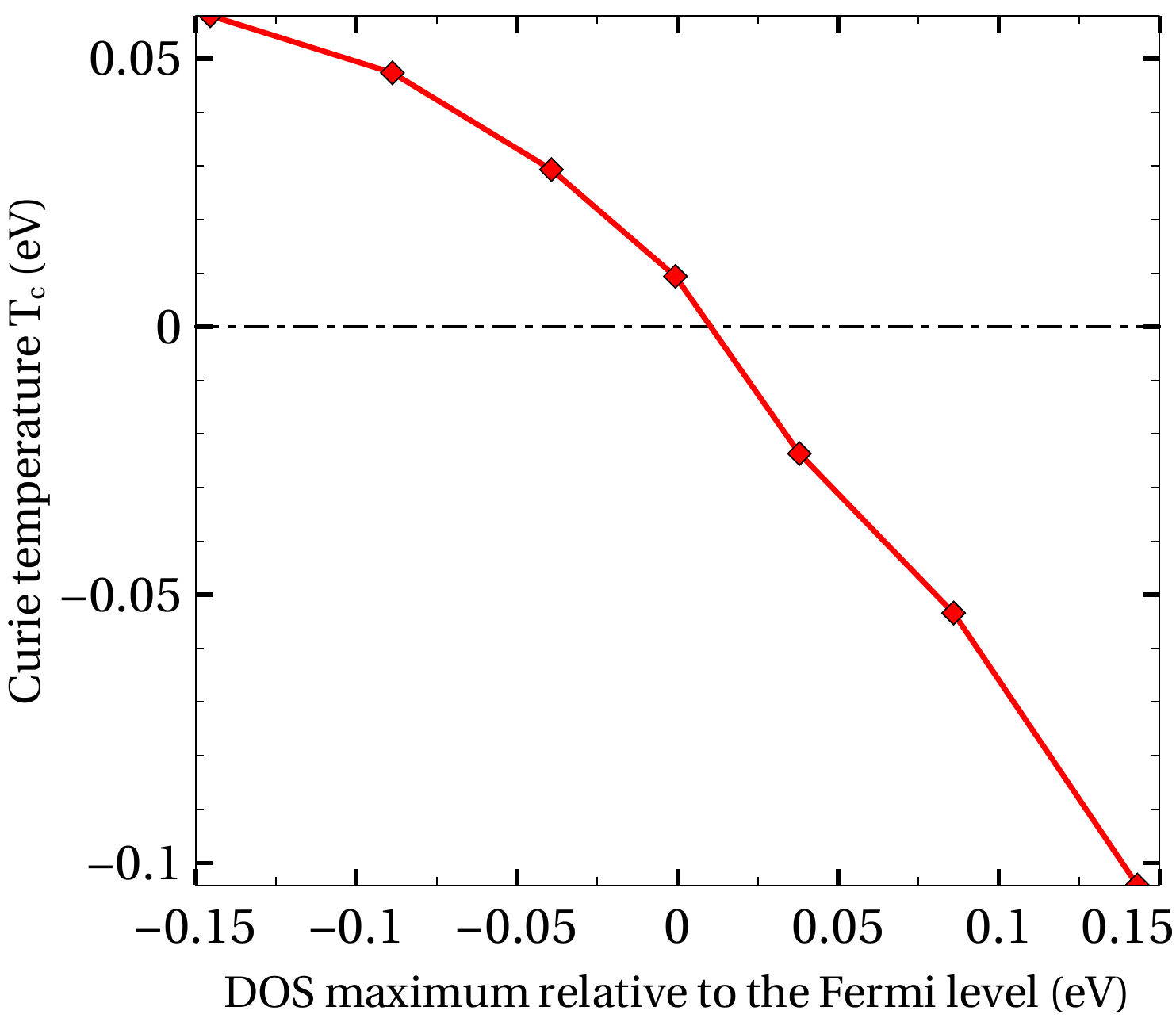}
\caption{\label{fig:peak_pos}(Color online) The dependence of Curie temperature $T_c$ on van Hove peak position with respect to the Fermi level of fully spin-polarized state calculated from Eqn.~\eqref{eqn:dispersion2D} with $t=0.264$eV, $t'/t$ changes from $-0.3\to0.3$. On-site interactions $U=6$eV and $J=1$eV. The dashed line marks the zero temperature.}
\end{figure} 

The importance of the position of the density of states peak was previously noticed in the context of the one-band Hubbard model.\cite{springerlink:10.1007/s002570050375,Euro.Phys.B1.301,PhysRevB.58.12749} However, in that case ferromagnetism was only found if the density of states peak is far below the Fermi level of the paramagnetic state. In the present case inspection of Fig.~\ref{fig:DOS_bulk} shows that it is only necessary for the density of states peak to be not too far above the Fermi level. We believe that the difference arises from  the ``double exchange'' physics of Hund's coupling in partially filled $d$ shells. The Hund's coupling favors high spin states, which means that hopping between two sites is optimal if the spins are parallel and is suppressed if they are not parallel. This  strongly favors ferromagnetism.  A natural question is how far above the Fermi level can the density of states peak be and still support ferromagnetism. For a reasonable range of $J$ ($\sim1$eV, see Fig.~\ref{fig:peak_pos}) we find that a good rule of thumb is that ferromagnetism occurs if the density of states peak lies at or below the Fermi level of the {\em fully polarized ferromagnetic state}. This answer is clearly not universal since model system studies\cite{Euro.Phys.B5.473,PhysRevB.83.125110,PhysRevB.80.235114} indicate that  increasing $J$ to very large values  favor ferromagnetism even if the density of states peak lies very high in energy.

We have also studied selected densities at the smaller correlation strengths $U=4$eV and $U=5$eV. Estimates of the resulting phase boundaries are shown as dashed and solid lines in Fig.~\ref{fig:phase_diagram}. We see that for intermediate $n$ the phase boundary is only weakly dependent on $U$ and $n$. For carrier concentrations near $n=2$ weakening the correlations to move the system out of the Mott phase greatly increases the range in which   ferromagnetism is observed. These calculations are the basis for  the three schematic phase diagrams shown in  Fig.~\ref{fig:schematic_pd}.

\section{The effect of oxygen bands\label{sec:pdmodel}}
In early transition metal oxides, the bands that cross the Fermi level are well separated from other bands and are of mainly transition metal $d$ character, arising from $t_{2g}$ symmetry orbitals. In many body calculations, it is common to focus only on these bands, which are treated as a multiband Hubbard model, while  all other bands are neglected.\cite{RevModPhys.70.1039,Pavarini04} However, it is generally believed that the fundamentally correct model of the transition metal oxides should be based on atomic-like $d$ orbitals coupled to weakly correlated oxygen $p$ states.\cite{PhysRevLett.55.418} In this section we examine the magnetic phase diagram of the vanadate system by applying the methods described in previous sections to  the ``$pd$'' model which describes this situation and comparing the ``$pd$'' model results to those from the ``$d$ only'' model.

The $pd$ model has two important energy scales: the on-site interaction $U$ which as before is the energy cost for changing $d$ occupancy and the charge transfer energy, in other words, the energy cost for  an electron to hop from a ligand to a transition metal atom. The charge transfer energy may be defined in different ways but the correct value is not well established, either from \textit{ab initio} considerations or from experiment.  The important points for our subsequent discussion  are that the physics depends sensitively on the charge transfer energy as well as on  $U$ and that for reasonable $U$ a metal-insulator transition may be driven at integer band filling by varying the charge-transfer energy.\cite{PhysRevLett.55.418} We will see that, as was found in the $d$ only model, the ferromagnetic phase boundary depends on whether the parameters are such as to place LaVO$_3$ on the metallic or insulating sides of the metal-insulator phase diagram.

\begin{figure}[h]
 \includegraphics[width=0.9\columnwidth]{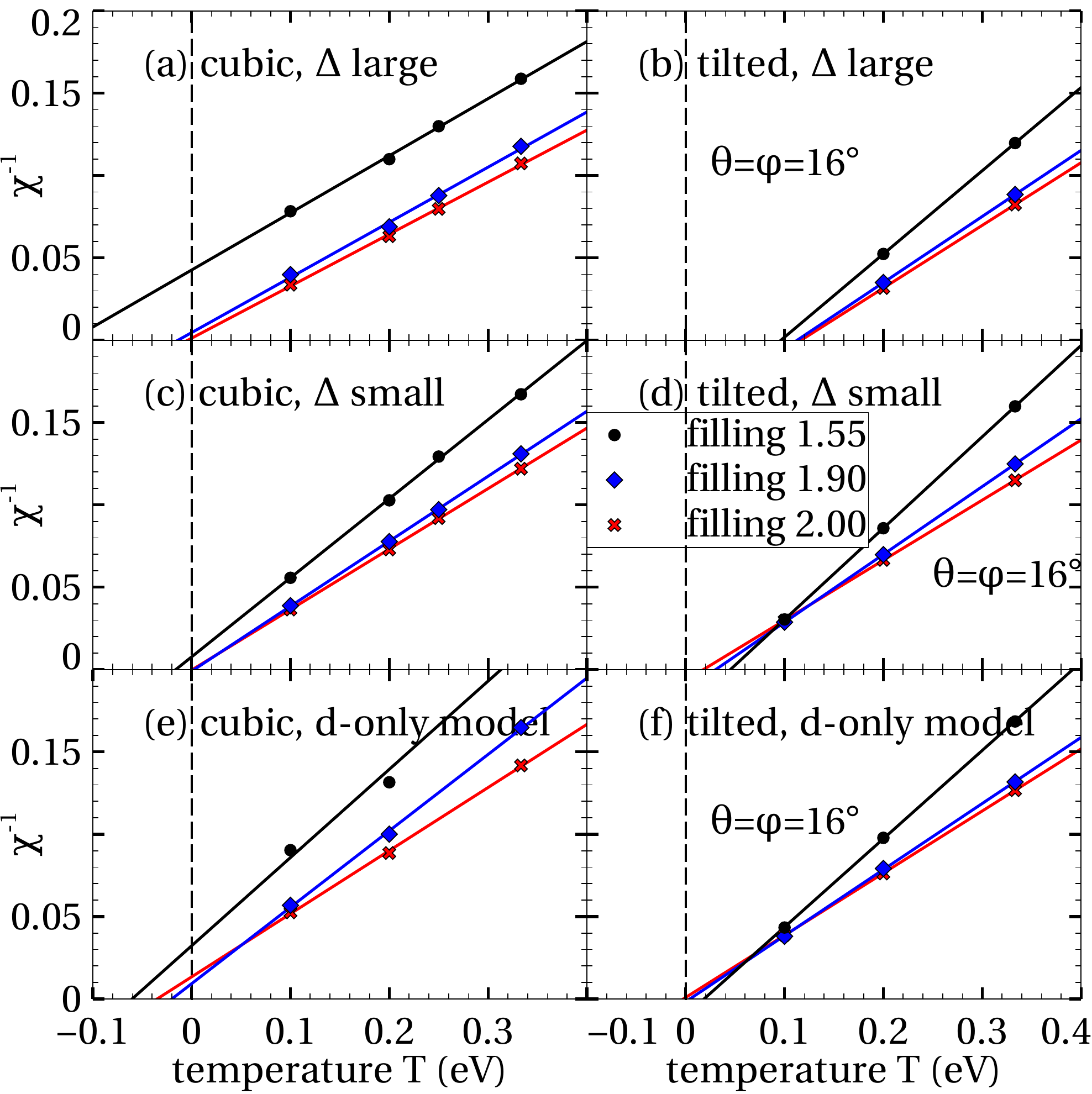}
\caption{\label{fig:Tc_trend_pdmodel}(Color online) Inverse susceptibility vs. temperature for three different filling values $1.55,1.90$ and $2.00$ for cubic (left column) and GdFeO$_3$-distorted LaVO$_3$ with tilt angle $\theta=\phi=16^\circ$ (right column). (a,b) $pd$ model with large $\Delta$ (approximately   that predicted by standard \cite{Lichtenstein98} double counting correction); (c,d) $pd$ model with small $\Delta$; (e,f) $d$ only model. On-site interactions $U=6\mathrm{eV}, J=1$eV. For cubic case, $\Delta_{large}=10.97\mathrm{eV},\Delta_{small}=2$eV. For tilted case, $\Delta_{large}=10\mathrm{eV},\Delta_{small}=0$eV. The vertical dashed lines mark zero temperature.}
\end{figure}

The Hamiltonian describing the $p$ and $d$ states of a perovskite transition metal oxide has the form\cite{PhysRevB.86.195136}
\begin{equation}
 H = H_d + H_p + H_{hyb},
\end{equation}
where $H_d = H_{kin}^d + H_{onsite}$ with $H_{kin}^d$ describing bare on-site energies and electron hoppings between $d$ orbitals and $H_{onsite}$ as in Eq.~\eqref{eqn:Honsite}; $H_p$ describes the dispersion of the oxygen $p$ bands in the absence of $p$ $d$ hybridization; $H_{hyb}$ is the hybridization between $d$ and $p$ orbitals. We take $H_{onsite}$ to have the usual rotationally invariant Slater-Kanamori form with  $U=6$eV and $J=1$eV.  The band term  $H_{kin} = H_{kin}^d+H_p+H_{hyb}$ has the same generic form as  Eq.~\eqref{eqn:Hkin} but now the space of  orbital indices $\alpha,\beta$ is expanded to label  both vanadium $d$ and oxygen $p$ orbitals. The corresponding $H_{band}^{\alpha\beta}(\mathbf{k})$ is generated using MLWF methods with a wide energy  window including both $d$ and $p$ bands (see the MLWF procedure in Sec.~\ref{sec:model} and the Appendix~\ref{app:band_wan}).  

However,\cite{Lichtenstein98,Kotliar06,Karolak10} the Hartree shift arising from $H_{onsite}$ means that the  $d$ level energy $\epsilon_d^0$ obtained from the MLWF procedure must be renormalized by a ``double counting correction'' which we denote by  $\Delta$: $\epsilon_d = \epsilon_d^0 - \Delta$.  We consider two values for $\Delta$: one, which we refer to as $\Delta_{large}\sim10$eV, is essentially the value obtained by applying the  standard\cite{Lichtenstein98} double counting correction to basic band theory and is such that LaVO$_3$ is predicted to be metallic, in contradiction to experiment.  The other value, which we refer to as $\Delta_{small}\sim0-2$eV, is such that the material is insulating at $n=2$ in agreement with experiment.  

The resulting  model is solved using single-site DMFT as described in Sec.~\ref{sec:model}, but with one important addition. The full $p$ $d$ manifold includes V $e_g$ orbitals. While the $e_g$ energy lies above the Fermi level, so that the $e_g$-derived antibonding bands are empty, some of the filled bands are $e_g$-oxygen bonding states which have a small but non-zero $e_g$ content. One must therefore solve a $5$-orbital impurity model. Performing this  calculation in full generality would be prohibitively  expensive. We therefore follow standard procedure and treat the $e_g$ orbital contribution to the impurity model in a Hartree approximation.

Fig.~\ref{fig:Tc_trend_pdmodel} shows representative computations of the inverse susceptibility for cubic and tilted systems at several carrier concentrations. The top row (panels a and b) displays $pd$ model  results obtained for the standard double counting correction (so that LaVO$_3$ is wrongly predicted to be a metal) while the middle rows (panels c and d) show results obtained if the double counting correction is tuned so that the calculation places LaVO$_3$ in the Mott/charge-transfer insulating regime of the phase diagram. The bottom two panels (e and f) present $d$ only model results for comparison. In the $d$ only model the $U$ value is such as to place the $n=2$ (LaVO$_3$) material in the Mott insulating region of the phase diagram). The transition temperature estimates obtained by linearly extrapolating the $\chi^{-1}$ curves to $0$ are given in Table \ref{table:Tc_value_pdmodel}.

\begin{table}[h]
 \begin{center}
\begin{tabular}{|l|c|c|c|}
\hline\hline
& $n=1.55$ & $n=1.90$ & $n=2.00$ \\
\hline
cubic, $\Delta_{large}=10.97$eV & -0.1230 & -0.0139  & -0.0040\\
tilted, $\Delta_{large}=10$eV   & 0.0963  & 0.1127  & 0.1167\\
cubic, $\Delta_{small}=2$eV     & -0.0161 & 0.0025  & 0.0007\\
tilted, $\Delta_{small}=0$eV    & 0.0452  & 0.0303  & 0.0178\\
cubic, $d$ only model           & -0.0602 & -0.0200 & -0.0349\\
tilted, $d$ only model          & 0.0185  & 0.0042  & -0.0025\\
cubic, $d$ only model, $U=4$eV  & -0.2347 & -0.1368 & -0.1250\\
tilted, $d$ only model, $U=4$eV & 0.0267  & 0.0345  & 0.0366\\
\hline\hline
 \end{tabular}
 \end{center}
\caption{\label{table:Tc_value_pdmodel} Values for Curie temperature $T_c$ (in eV) for each case considered in Fig.~\ref{fig:Tc_trend_pdmodel} together with results for $U=4$eV for three different fillings $n=1.55,1.90$ and $2.00$.  All computations are for $J=1$eV; except where indicated, $U=6$eV is used.}
\end{table} 

Examination of the results in Table.~\ref{table:Tc_value_pdmodel} shows that  the qualitative trends are the same in the $pd$ and $d$ only model calculations. In particular, in both models increasing the tilt angle increases the tendency towards ferromagnetism.  However, significant differences are visible; in particular  the  $pd$ model has a significantly greater tendency to ferromagnetism than does the $d$ only model and (especially in the small-$\Delta$ case)  the differences are more pronounced for the cubic than for the tilted structure. 

We believe that there are two main origins for the differences. First, in the small $\Delta$ cubic system case, the change in the charge transfer energy relative to band theory affects the density of states, moving the peaks closer to the Fermi level while for the tilted structure the shift in charge transfer energy does not change the peak positions as much  (see Appendix C).  Second, and perhaps more important, the $e_g$ state occupancy arising from the $p-d$ bonding bands (omitted in the $d$ only model) increases the effective moment on the $d$ site, thereby enhancing the tendency towards magnetism. This effect is more pronounced in the larger $\Delta$ (smaller charge transfer energy) case, because the $p$ $d$ mixing is larger. 

\begin{figure}[h]
 \includegraphics[width=\columnwidth]{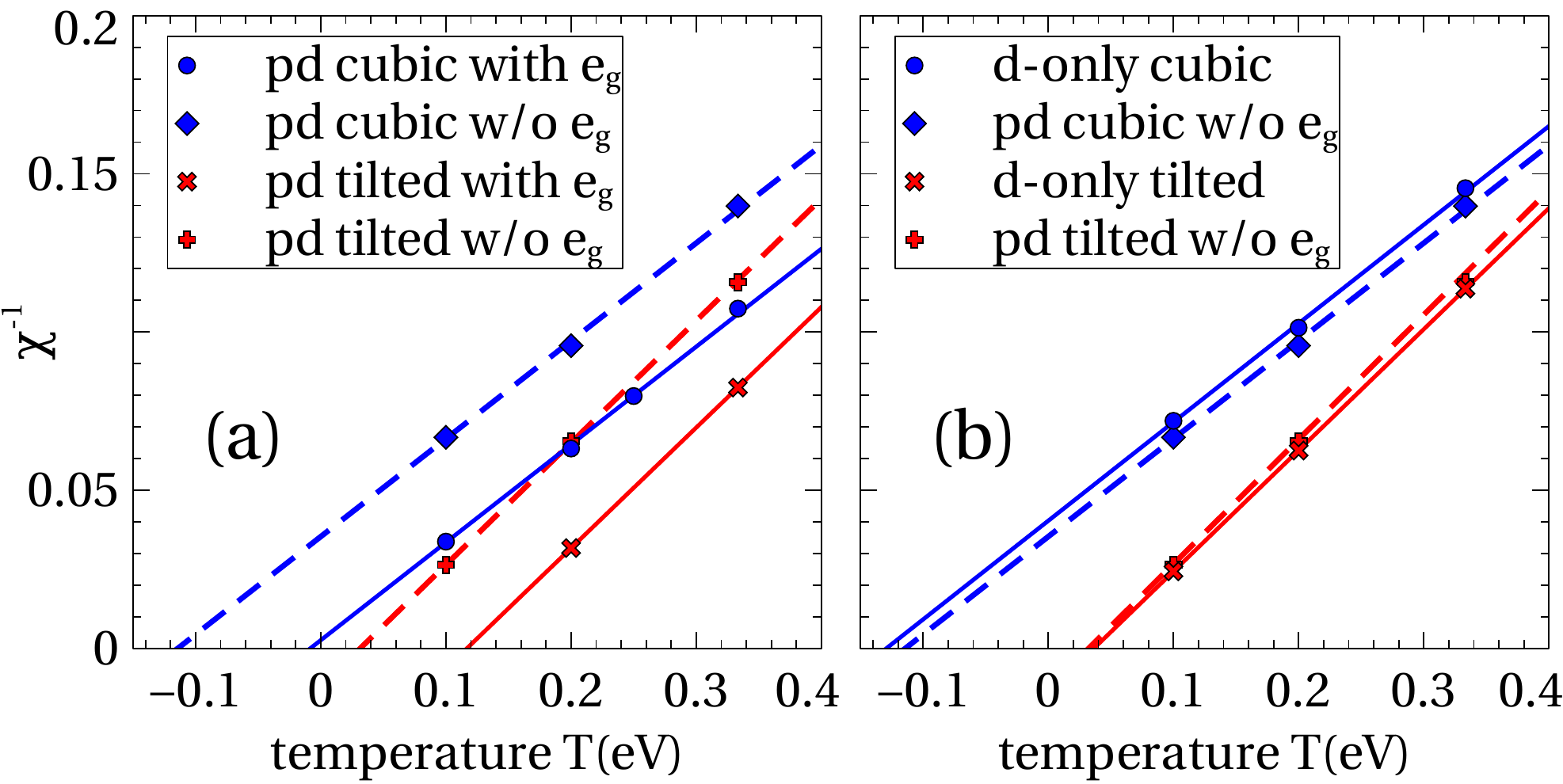}
\caption{\label{fig:eg_contrib}(Color online) Inverse susceptibility vs. temperature  for cubic (blue color) and $16^\circ$-tilted (red color) structures at filling $n=2.0$. (a) Comparison for $pd$ model with (solid lines) and without (dashed lines) $e_g$ band spin polarization. (b) Comparison between $d$ only model (solid lines) and $pd$ model without $e_g$ band spin polarization (dashed lines). Parameters for calculations with $pd$ model are the same as in Fig.~\ref{fig:Tc_trend_pdmodel}a,b. Calculations with $d$ only model use $U=4\mathrm{eV}<U_c, J=1$eV for both structures.}
\end{figure}

To demonstrate this point we present in  Fig.~\ref{fig:eg_contrib}a,b calculations of the inverse susceptibility under different conditions. The curves in the left panel compare calculations in which the $e_g$ occupancy is frozen at the spin unpolarized values (dashed lines) and calculations in which the $e_g$ orbitals are treated within the Hartree-Fock approximation as described above. We see that the feedback from the polarization of the $e_g$ orbitals makes a significant contribution to the transition temperature. The right panels show that calculations performed in the ``frozen $e_g$'' model agree reasonably well with the corresponding calculations in the $d$ only model for $U=4eV<U_c$. Further analysis is given in Appendix~D.

\section{Conclusions\label{sec:conclusions}}

In this paper we have investigated the conditions under which ferromagnetism might be observed in bulk solid solutions derived from LaVO$_3$. Our theoretical studies used realistic band structures derived from maximally-localized Wannier function fits to density functional (typically PWscf/GGA) band structure calculations, along with the single-site DMFT approximation which contains the local physics of partially filled orbitally degenerate $d$ orbitals with strong Hund's coupling.  Our main theoretical finding is that for physically reasonable values of the Hund's coupling ferromagnetism is driven mainly by  density of states effects, being favored by a density of states peak lying not too high in energy. The dependence on correlation strength and carrier concentration is relatively weak except that if the correlation strength is large enough to drive a Mott transition to an antiferromagnetic Mott/charge transfer insulator, ferromagnetism is suppressed in the vicinity of the insulating state. 

An association between ferromagnetism and the density of states was previously noticed in studies of the one band Hubbard model.\cite{Vollhardt.Review,springerlink:10.1007/s002570050375,PhysRevB.80.235114} However, there is a significant difference between ferromagnetism in the model studied here and that found in the  single-band Hubbard model. In the single-band Hubbard model, ferromagnetism only occurs when the DOS peaks are located far below the chemical potential, and indeed very close to the lower band edge.\cite{Vollhardt.Review} In the orbitally degenerate, Hund's coupled systems studied here, the requirements on the position of the density of states peak are substantially relaxed. We find that for physically reasonable correlation strengths, ferromagnetism can occur as long as  the lowest density of states peak is at or below the Fermi level of the fully polarized ferromagnetic state. The importance of orbital degeneracy and Hund's coupling has been previously noted in studies of models with a semicircular (Bethe-lattice) density of states\cite{Euro.Phys.B5.473,PhysRevB.83.125110,PhysRevB.80.235114} and in the periodic Anderson model\cite{PhysRevB.78.205118} but the important role played (for reasonable values of $J$) by the near-Fermi-surface density of states peaks seems not to have previously been noticed. 

In the vanadate systems, the density of states is controlled by the amplitude of the GdFeO$_3$ distortion away from the ideal cubic perovskite structure. We find that increasing the distortion amplitude favors ferromagnetism. Indeed, if the $U$ interaction were set to the unphysically small value $U=4$eV so that LaVO$_3$ were not a Mott insulator, then the observed GdFeO$_3$ distortion would be large enough to put the material at or very near to the ferromagnetic phase boundary. One may then ask why doping does not induce ferromagnetism in  bulk solid solutions such as La$_{1-x}$Sr$_x$VO$_3$. We believe that the answer is that in the physical system, the distortion amplitude and the carrier concentrations are both determined by the La/Sr ratio in such a way that, as $x$ is varied, the distortion amplitude decreases so that the system follows a path in parameter space which remains outside the ferromagnetic region (solid line in  Fig.~\ref{fig:phase_diagram}).

We remark that while the general experience has been that the single-site DMFT approximation provides a good qualitative representation of the physics, in particular of trends as material parameters are varied,\cite{Pavarini04,Kotliar04,Kotliar06,Held07} its quantitative accuracy in producing magnetic phase boundaries has not been established. It is possible that single-site DMFT underestimates magnetic order. The reports of ferromagnetism in vanadate superlattices\cite{PhysRevB.80.241102} where both the doping and the tilt angle may be locally large, suggests that the true ferromagnetic phase boundary may be shifted to lower tilt angle but still above the bulk LaVO$_3$ $\theta(n)$ curve. Increases in computer power and improvements in algorithms may soon make cluster dynamical mean field studies of realistic systems with orbital degeneracy and Hund's coupling feasible, and it will be important to use these methods to assess the reliability of single-site DMFT predictions. 

The second point of materials theory is the issue of what is the relevant theoretical model. Consistent with most theoretical literature on early transition metal oxides,\cite{RevModPhys.70.1039,Pavarini04,Kotliar04}  we focused mainly on a $d$ only ``multiband Hubbard model'' approach where the electrons in the near Fermi surface $d$ derived bands were correlated by local $U, J$ interactions. The results are in reasonable accord with experiment. It has been argued \cite{PhysRevLett.55.418,Kotliar06,Held07} that a more generally valid description may be obtained from a model in which correlations are applied to atomic-like $d$ orbitals which are hybridized to the other orbitals, in particular to the oxygen $p$ orbitals. In this ``$pd$'' model approach, a key parameter is the renormalized $d$ level energy, which is shifted from the band theory value by a ``double counting correction'' for which there is no generally agreed value. If the standard expression\cite{0953-8984-9-4-002} for the double counting correction is  used, in the single-site DMFT approximation LaVO$_3$ is found to be a ferromagnetic metal rather than an antiferromagnetic insulator and indeed the solid solution La$_{1-x}$Sr$_x$VO$_3$ is predicted to be ferromagnetic for a wide range of $x$, in reasonable correspondence to the small $U$ calculation in the $d$ only model. If the double counting contribution is shifted such that  the model for LaVO$_3$ is in its Mott insulating regime, then the resulting phase diagram is very similar to that shown in Fig.~\ref{fig:phase_diagram} for the $d$ only model. Quantitative differences arise from the changes in density of states arising from the  large renormalization of $\epsilon_d$ required to make the $n=2$ case a Mott insulator and from the contribution of the (small but non-negligible)  $e_g$ occupancy arising from the the $p$-$d$ bonding states.   Determining the appropriate theoretical approach for the early transition metal oxide is an important open question.

Our findings suggest several routes to creating ferromagnetism in artificially fabricated systems. The general aim should be to manipulate the band structure so as to move density of states peaks close to the Fermi level. Inducing octahedral tilts by compressive strain is an important route. Further, in early transition metal oxides,  increasing the $p$ $d$ hybridization clearly increases the effective $d$ moment and this provides a self-consistent amplification of the Curie temperature. Therefore manipulation of the $p$ $d$ energy difference can be used to control $T_c$. 

We also observe that  some transition metal oxides such as the Ru-based perovskites and Ruddlesden Popper materials Sr$_{n+1}$Ru$_n$O$_{3n+1}$ involve holes in the $t_{2g}$ bands and in a qualitative sense may be thought of as the particle-hole transforms of the models studied here. In this case for the physical sign of $t'$ the van Hove peaks are on the other side of the Fermi level, suggesting that the theoretical models will be more likely to exhibit ferromagnetism. However,  increasing the tilt angles shifts the peaks in the wrong direction, reducing the tendency to magnetism. A more detailed investigation of this physics is in progress.

\section*{ACKNOWLEDGEMENTS}
We thank U. L\"uders, J. Okamoto and C. Marianetti for helpful conversations. We acknowledge support from DOE-ER046169. HTD acknowledges partial support from the Vietnam Education Foundation (VEF). A portion of this research was conducted at the Center for Nanophase Materials Sciences, which is sponsored at Oak Ridge National Laboratory by the Scientific User Facilities Division, Office of Basic Energy Sciences, U.S. Department of Energy. We used the code for CT-HYB solver\cite{PhysRevLett.97.076405} written by P. Werner and E. Gull, based on the  ALPS library.\cite{Albuquerque20071187}

\appendix

\section{\texorpdfstring{$\theta(x)$}{theta} Curve for \texorpdfstring{La$_{1-x}$Sr$_x$VO$_3$}{La1-xSrxVO3}}\label{app:thetax}
The relation between tilt angle $\theta$ and the doping level $x$ is obtained based on the fact that the (pseudo)cubic lattice constant $a_p$ of La$_{1-x}$Sr$_x$VO$_3$ is linearly dependent on $x$.\cite{PhysRevB.46.10973} Bulk SrVO$_3$ has $a_p^{SVO} = 3.84$\AA,\cite{Rey1990101} bulk LaVO$_3$ has $a_p^{LVO} = 3.92$\AA,\cite{Bordet1993253} hence for doped material, $a_p(x) = a_p^{SVO} x + a_p^{LVO} (1-x)$. 

We further assume that the effective V-O distance $d$ also has similar relation, $d(x) = d_{SVO}x + d_{LVO}(1-x)$. Bulk SrVO$_3$ is cubic, thus $d_{SVO} = \dfrac{a_p^{SVO}}{2}=1.92$\AA; while $d_{LVO} = 2.00$\AA.\cite{Bordet1993253} The angle $\theta$ is calculated from $a_p = 2 d\cos\theta$, and with $n=2-x$ we obtain the $\theta(n)$ curve shown in  Fig.~\ref{fig:phase_diagram}.

\section{Band Calculations and Wannier Fitting}\label{app:band_wan}

Our procedure for obtaining the band theoretic contribution to the local Green's functions is as follows. First, given a rotation angle $\phi$ and tilt angle $\theta$ we construct the 4-unit-cell structure corresponding to a GdFeO$_3$-type distortion of the cubic perovskite. The PWscf code of QUANTUM ESPRESSO\cite{QE-2009,QEPseudo} is then  used to calculate the band structure for that system. Finally, we use the Wannier90 implementation\cite{Mostofi2008685} of the MLWF procedure with an appropriate energy window to generate our $\hat{H}_{band}(\mathbf{k})$. 

For the band structure calculations with PWscf, we used a cutoff energy $E_{cutoff}=30\mathrm{Ry}\approx408$eV and a $10\times10\times10$ Monkhorst-Pack $\mathbf{k}$-mesh.  The MLWF procedure involves the choice of both an overall energy window and a ``frozen'' window within which the MLWF bands are forced to coincide with the DFT bands. For MLWF fitting with Wannier90, we set the overall energy window for the $d$ bands to run from $-1.5$eV to $6$eV with respect to the chemical potential. For larger octahedral rotations, the $e_g$ and $t_{2g}$ bands are well separated and choosing the frozen energy window to run from $-1.5$eV to $1.5$eV represents the $t_{2g}$ bands well. For smaller rotations, band overlap becomes  important and the upper cutoff of the frozen energy window must be reduced to avoid overlap with the $e_g$ bands; this means that the high-lying (unoccupied) part of the $t_{2g}$ bands is not perfectly represented, but these issues do not affect our main results, which concern the location of the 
ferromagnetic phase. If oxygen bands are included in the calculation, the energy window is enlarged from $-10$ to $6$eV, while the range $-10\to 1.5$eV is set as the frozen energy window.

\section{Charge Transfer Model}
\begin{figure}[h]
 \includegraphics[width=0.9\columnwidth]{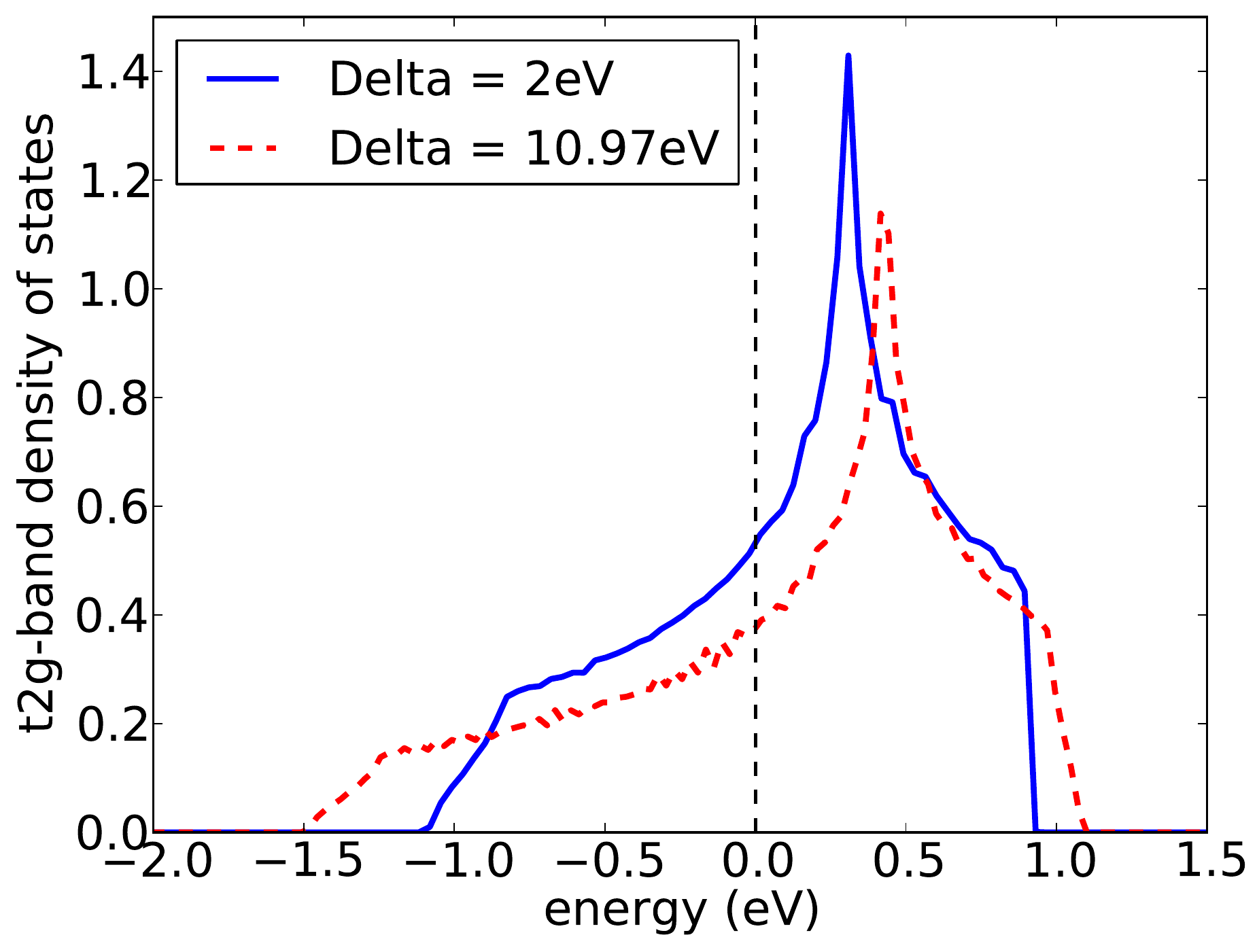}
\caption{\label{fig:pd_dos}(Color online) Hartree-Fock orbital/spin unpolarized $t_{2g}$ band density of states with $U=6$eV and $J=1$eV for two cases of cubic structure used in Sec.~\ref{sec:pdmodel}: $\Delta_{small}=2$eV and $\Delta_{large}=10.97$eV. The dashed vertical line marks the Fermi level.}
\end{figure}

In the charge transfer ($pd$) model, the value of the charge transfer energy $\Delta$ affects the ferromagnetic phase boundary and also changes the band structure. In Fig.~\ref{fig:pd_dos} we demonstrate the effect of varying  $\Delta$ on the band structure of cubic LaVO$_3$. 

We use the same parameters as in Sec.~\ref{sec:pdmodel} for the cubic structure and employ a Hartree-Fock approximation to derive the spin/orbital unpolarized DOS. Fig.~\ref{fig:pd_dos} shows such DOS for $\Delta=2$eV and $\Delta=10.97$eV. Within Hartree-Fock calculation, oxygen $p$ bands are located at the same positions as in corresponding DMFT results, so the $p$ $d$ hybridization may be expected to be similar in the two calculations. We therefore consider the Hartree-Fock DOS as the  ``non-interacting'' DOS for the $pd$ model, including charge transfer effects.

The two curves shown in Fig.~\ref{fig:pd_dos} demonstrate that changing $\Delta$ changes both the position of the van Hove peaks relative to the Fermi level and the bandwidth. For $\Delta_{large}=10.97$eV, the bandwidth is $W=2.58$eV and the ratio between next nearest neighbor and nearest neighbor hoppings is $t'/t=0.23$. For $\Delta_{small}=2$eV, the bandwidth $W$ of the antibonding band  is smaller ($W=2.03$eV) and the van Hove peak moves closer to the Fermi level ($t'/t=0.18$). These two factors act to make the model more ferromagnetic.

\section{Contribution of \texorpdfstring{$e_g$}{e\_g} Bands}
In present-day applications of  the  DFT+DMFT method to  transition metal oxides, atomic-like $d$ orbitals are defined via a projector or Wannier construction and are coupled to other bands (mainly oxygen $p$, in practice).  The resulting theory differs from the multiband Hubbard models often used \cite{Pavarini04} in two ways. First, the explicit inclusion of oxygen orbitals means that charge transfer physics in the sense of Ref.~\onlinecite{PhysRevLett.55.418} may play a role. Second, the full $d$ manifold is included, in contrast to the multiband Hubbard approach where only the frontier orbitals (in the present case, the $t_{2g}$ states) are treated.   This latter difference is important for the vanadates because even though the $e_g$ orbitals are notionally empty, hybridization with the oxygen bands leads to a non-vanishing density of $e_g$ electrons. 

A fully interacting treatment of the entire $d$ manifold is too expensive to be computationally practicable. In this paper we take the virtual occupation of the $e_g$ orbitals into account via a Hartree approximation. In each iteration of the DMFT self consistent calculation, $e_g$ occupancies obtained from the previous iteration are used to calculate the $e_g$ feedback to the $t_{2g}$ bands via the Hartree approximation to the inter-orbital self energy:
\begin{equation}
 S^H_\sigma = \sum_{i\text{ in }e_g\text{ bands}}\left[(U-2J)n_{i\bar{\sigma}} + (U-3J)n_{i\sigma}\right].
\end{equation}
The two panels of Fig.~\ref{fig:eg_contrib} show that it plays an important role in the magnetic phase diagram. In what follows we give analytic Stoner-style arguments explaining these results.

To estimate the contribution of $e_g$ bands to the polarization, we first observe that although the $e_g$ orbitals are notionally empty, hybridization with the oxygen states means that the  $e_g$ occupancy is non-vanishing; roughly if the $p$ $e_g$ energy difference is $\Delta_e$ and the $p-d$ hopping is $t_{pd}$  then 

\begin{equation}
n_{e_g}\sim \frac{t_{pd}}{\Delta_e}.
\label{neg}
\end{equation}
A perturbation to the $e_g$ energy of magnitude $h$ thus leads to a change in occupancy
\begin{equation}
\delta n_{e_g}\sim \frac{t_{pd}h}{\Delta_e^2},
\label{neg2}
\end{equation}
so that we expect a  susceptibility  given by
\begin{equation}
\chi_{e_g}\sim \frac{t_{pd}}{\Delta_e^2}.
\label{chieg}
\end{equation} 
The absence of $e_g$ character at the Fermi level and the fact that the $e_g-p$ energy difference ($\Delta_e$) is larger than that of the $t_{2g}-p$ energy difference suggests that $\chi_{eg}$ will be smaller than the band susceptibility, but not by orders of magnitude.

We will assume that the intra-$t_{2g}$ contributions lead to a $\chi_{t_{2g}}=A/(T-T_0)$ which would diverge at some temperature $T_0$. In the presence of a magnetic field this $\chi_{t_{2g}}$ leads to a polarization of the $t_{2g}$ orbitals which, via the $J$ interaction, affects the $e_g$-oxygen $p$ energy difference and thereby leads to a polarization of the $e_g$ orbitals which in turn feeds back on the $t_{2g}$ states. Mathematically, we distinguish the $e_g$ and $t_{2g}$ orbitals and write $m_{e_g,t_{2g}}$ as the magnetization of one orbital
\begin{equation}
 \left\{
\begin{array}{c}
 m_{t_{2g}} = \chi_{t2g}(h +  2J m_{e_g}),\\
 m_{e_g} = \chi_{eg}(3J m_{t_{2g}} + J m_{e_g}).
\end{array} 
\right.
\end{equation}
Solving these equations gives
\begin{equation}
\begin{split}
\dfrac{h}{m_{t2g}}&=\chi_{t2g}^{-1}-\dfrac{6J^2}{\left(\chi_{eg}\right)^{-1} -J} \\
&\approx A^{-1}(T-T_0) - 6J^2\chi_{eg}.
\end{split}
\end{equation}

Thus the effect of the $e_g$ component of the bonding bands is to shift the Curie temperature  by $\Delta T\approx 6J^2A\chi_{eg}$.  From Fig.~\ref{fig:eg_contrib}, we estimate $A\approx 3$ so even though although $\chi_{eg}$ is likely to be small, the overall effect may be non-negligble.

\end{document}